%% file: tr.tex
\documentclass[11pt]{article}

\usepackage{mathptmx}
\usepackage{authblk}
\usepackage{helvet}
\usepackage{courier}
\usepackage{graphicx}
\usepackage{multicol}
\usepackage[bottom]{footmisc}
\usepackage{numprint}
\usepackage{doi}

\usepackage[dvipsnames]{xcolor}
\usepackage{url}

\usepackage[ruled,vlined,linesnumbered]{algorithm2e}
 \DontPrintSemicolon
 \SetAlFnt{\small}
 
 \SetAlgoCaptionLayout{sppcapsty}
 \newcommand{\sppcommentsty}[1]{\textcolor{black!50}{\sffamily #1}}
 \SetCommentSty{sppcommentsty}
 
 \SetDataSty{sppdatasty}
 
 \SetFuncSty{sppkwfunctiondatasty}

\usepackage{listings}
\usepackage{amsmath}
\usepackage{amsfonts}
\usepackage{amssymb}

\usepackage{hyperref}  
\usepackage[capitalise]{cleveref}

\usepackage[group-separator={\,},binary-units=true,exponent-product = \cdot, output-product = \cdot]{siunitx}

\usepackage{tikz}
\usetikzlibrary{arrows}
\usetikzlibrary{backgrounds}
\usetikzlibrary{shapes.geometric}
\usetikzlibrary{shapes.multipart}
\usetikzlibrary{arrows.meta} 
\usetikzlibrary{decorations.pathmorphing}
\usetikzlibrary{calc}
\usetikzlibrary{shapes}
\usetikzlibrary{patterns}
\usetikzlibrary{decorations.pathreplacing}

\graphicspath{chapter/figures/}

\usepackage{etoolbox}
\usepackage{ifthen}
\usepackage{xspace}
\usepackage{lipsum}
\usepackage[sort&compress,numbers]{natbib}
\usepackage{fullpage}

\newcommand{\Oh}[1]{\mathcal O(#1)}

\newcommand{\etal}{~et~al.}
\newcommand{\eg}{e.g.,}
 
\newcommand{\ie}{i.e.,}

\date{}
\newenvironment{acknowledgement}
    {
          \ \\  \noindent \textbf{Acknowledgements}. \\
    }
    { 
    }
\title{Recent Advances in Practical Data Reduction} 

\author[1]{Faisal Abu-Khzam}
\affil[1]{Lebanese American University, Lebanon}
\author[2]{Sebastian Lamm}
\affil[2]{Karlsruhe Institute of Technologie, Germany}
\author[3]{Matthias Mnich}
\affil[3]{Hamburg University of Technology, Institute for Algorithms and Complexity, Germany}
\author[4]{Alexander Noe}
\affil[4]{University of Vienna, Austria}
\author[5]{Christian Schulz}
\affil[5]{Heidelberg University, Germany}
\author[6]{Darren Strash}
\affil[7]{Hamilton College, USA}

\begin{document}%

	\maketitle
\begin{abstract}
Over the last two decades, significant advances have been made in the design and analysis of fixed-parameter algorithms for a wide variety of graph-theoretic problems.  This has resulted in an algorithmic toolbox that is by now well-established.  However, these theoretical algorithmic ideas have received very little attention from the practical perspective.  We survey recent trends in data reduction engineering results for selected problems. Moreover, we describe concrete techniques that may be useful for future implementations in the area and  give open problems and research questions.
\end{abstract}

\input{chapter/content.tex}

	\bibliographystyle{abbrvnat}
  \small
	\bibliography{chapter/references.bib}
\end{document}

%% file: chapter/content.tex
\newtheorem{openproblem}{Open Problem}
\newcommand{\csch}[1]{{\color{red} [Christian asks:]{#1}}}
\newcommand{\slamm}[1]{{\color{purple} [Sebastian asks:]{#1}}}
\newcommand{\dstrash}[1]{{\color{blue} [Darren asks:]{#1}}}
\def\MdR{\ensuremath{\mathbb{R}}}
\def\MdN{\ensuremath{\mathbb{N}}}
\newcommand{\wgt}{\mathcal{W}}
\newcommand{\quasikernel}{quasi kernel}

\thispagestyle{empty}
\section{Introduction}
Many important real-world optimization problems are NP-hard: it is believed that no polynomial-time algorithm exists that always finds an optimal solution.  However, many NP-hard problems have been shown to be fixed-parameter tractable (FPT): large inputs can be solved efficiently and provably optimally, as long as some problem parameter is small.  Over the last two decades, significant advances have been made in the design and analysis of fixed-parameter algorithms for a wide variety of graph-theoretic problems. 
This has resulted in an algorithmic toolbox that are by now well-established. 
However, these theoretical algorithmic ideas have received very little attention from the practical perspective. 
Until recently, few fixed-parameter algorithms are implemented and tested on real data sets, and their practical potential is far from understood.  
Traditionally, algorithms are designed using simple models of problems and machines.
In turn, important results are provable, such as performance guarantees for all possible inputs.
This often yields elegant solutions being adaptable to many
applications with predictable performance for previously unknown inputs. 

In algorithm theory, however, taking up and implementing an algorithm
is part of the application development. 
Unfortunately, this mode of transferring results is a slow process and sometimes the theoretically best algorithms perform poorly in practice. Hence, practitioners often do not read research papers from the algorithms theory community. This causes a growing gap between theory
and practice: Realistic hardware with its parallelism, memory hierarchies etc.~is
diverging from traditional machine models. 
This gap is also partially due to the fact that the research community working on algorithmic problems is fairly separated. One the one hand, there are ``hard core'' algorithms researchers that are focused mainly on theoretical work and rarely  participate in conferences in application areas. On the other hand, researchers of application areas publish their work in conferences and journals of their respective field, and often do not visit theory conferences. 
In contrast to algorithm theory, algorithm engineering uses an innovation cycle where
algorithm design based on realistic models, theoretical analysis, efficient
implementation, and careful experimental evaluation using real-world inputs
closes gaps between theory and practice and leads to improved application codes
and reusable software libraries (see {\small\url{www.algorithm-engineering.de}}). This yields results that practitioners can rely on for their specific application. 

 On the one hand, experimental results can trigger new theoretical questions and suggest new properties of inputs that are relevant parameters to use in theoretical analysis. On the other hand, the rich toolbox of parameterized algorithm theory offers a rich set of algorithmic ideas that are challenging to implement and engineer in practical settings. By applying techniques from fixed-parameter algorithms in nontrivial ways, algorithms can be obtained that perform surprisingly well on real-world instances for NP-hard problems. The viability of this approach has been demonstrated in recent years through the Parameterized Algorithms and Computational Experiments Challenge (PACE)~\cite{DBLP:conf/iwpec/BonnetS18,DBLP:conf/iwpec/DellHJKKR16,dell2017pace,DBLP:conf/iwpec/DzulfikarFH19}, in which teams compete to solve real-world inputs using ideas from parameterized algorithm design. Many researchers from all over the world have participated in that challenge. Moreover, the viability of this approach has recently been demonstrated by a wide range of papers. As the engineering part in the area has recently gained some momentum, we survey recent results and techniques that started to bridge the gap between theory and practice that is currently observed in the area.

\emph{Theoretical Context.}
All known exact and deterministic algorithms that solve NP-hard problems require time that is at least super-polynomial in the total size of the input. However, some problems can be solved by algorithms that run in time which is exponential only in the size of a fixed parameter while polynomial in the size of the input; those are called \emph{fixed-parameter algorithms}.
Here, the parameterized problem can be solved efficiently for small values of the fixed parameter.
Formally, a parameterized problem is a language $L \subseteq \Sigma^* \times \MdN$, where $\Sigma$ is a finite alphabet. The second component is called the parameter of the problem.
A parameterized problem $L$ is \emph{fixed-parameter tractable} if the question $(x,k) \in L$ can be decided by an algorithm in running time $f(k)\cdot |x| ^ {\Oh{1}}$, where $f$ is a computable function depending on $k$ only. The corresponding complexity class is called FPT. 

Fixed-parameter tractability is closely related to kernelization.
In kernelization, we apply data reductions to the input graph~$G$ until it cannot be reduced further, producing~a~\emph{kernel}~$\mathcal{K}$.
Rules that are used to reduce the graph while retaining the ability to compute an optimal solution are called \emph{data reduction rules}, or simply \emph{reductions}.
Given an binary encoded instance $(x,k) \in \{0,1\} ^* \times \MdN$ of some parameterized problem $L$, a \emph{kernelization} for $L$ produces an instance $(x',k')$ in polynomial time that satisfies: $(x',k') \in L \Leftrightarrow (x,k) \in L$ and $|x'| + k' \leq f(k)$ where $f$ is a computable function. Note that $f$ only depends on the problem parameter~$k$. So roughly speaking, kernelization can be thought of as a preprocessing routine that reduces a given problem instance to its ``most difficult part''. The function $f$ measures the kernel size. If $f(k) = \Oh{k^c}$ for some constant $c$ then the kernel is called polynomial kernel, and we say the problem admits a polynomial kernel.

\emph{Organization.} The rest of the paper is organized as follows. We first survey recent data reduction engineering results for selected NP-hard problems, and then for problems in P. We then describe concrete techniques that may be useful for future implementations in the area. Lastly, we give open problems and research questions.

\section{Recent Advances for NP-Hard Problems}

\subsection{Maximum Independent Set and Minimum Vertex Cover}
Given an undirected graph $G=(V,E)$, the goal of the \emph{maximum independent set} (MIS) problem is to compute a set of vertices $I \subseteq V$ such that (1) no two vertices in~$I$ are adjacent to one another, (2) the set $I$ has maximum cardinality among all such sets.
The complement of an independent set $I$ is called a \emph{vertex cover}~$V \setminus I$.
The MIS problem and the complementary problem of finding a minimum vertex cover (MVC) are well-studied NP-hard optimization problems~\cite{garey1990guide} that attract both researchers and practitioners alike.
Furthermore, there is no polynomial time approximation algorithm for MIS that can provide an $\Oh{n^{1-\epsilon}}$ guarantee for any constant $\epsilon >0$, unless P$=$NP~\cite{zuckerman2006linear}.
Finally, MIS is $W[1]$-hard~\cite{downey1999parameterized} when parameterized by solution size $k$.
This makes it unlikely that the problem is fixed-parameter tractable in $k$~\cite{downey1999parameterized}.
On the other hand, MVC is fixed-parameter tractable in solution size~$k$~\cite{downey1999parameterized}.

\subsubsection{Exact Approaches}
In recent years, the bridge between theoretically efficient algorithms and their practical applicability has been significantly reduced.
In particular, the branch-and-reduce paradigm, \ie~branching algorithms that use a wide variety of reduction rules, have been (1) shown to achieve theoretical running times that are among the best for both MIS and MVC~\cite{fomin-2009,xiao-2017}, and (2) are able to solve large real-world networks in practice~\cite{akiba-tcs-2016}.
However, most often the approaches used in practice only use a small subset of the reduction rules that have been proposed to achieve good theoretical running times.

Abu-Khzam\etal~\cite{abu2004kernelization} introduced and analyzed the crown reduction rule (and the usage of data reduction rules in this context in practice). Even though the crown rule is not as powerful as the linear programming (LP)-based rule~\cite{nemhauser-1975} when considering the worst-case size of the resulting kernel, they experimentally verified that it often performs as well as the LP-based rule and is significantly faster in many cases.
Furthermore, they show that the LP-based rule is most useful for fairly sparse graphs and should be avoided for dense graphs, as it yields little to no reduction in size.

Later, Akiba and Iwata~\cite{akiba-tcs-2016} were the first to show the practicality of the branch-and-reduce paradigm for MVC (and MIS) compared to other state-of-the-art approaches like branch-and-bound and branch-and-cut.
Their algorithm uses a wide spectrum of reduction rules that form the foundation of many subsequent work.
This includes both conceptually simple reduction rules like degree-1 and degree-2 vertex folding~\cite{fomin-2009}, as well as more complicated but practically significant rules like unconfined~\cite{xiao2013confining} and an LP-based rule~\cite{iwata-2014,nemhauser-1975}.
Many of these reduction rules work by removing vertices that are part of some MIS.
For example, in the degree-1 reduction rule one removes vertices $v$ of degree one (and their neighbors), as they are always in at least one MIS.
To see this, note that $v$ or its neighbor $w$ must be in some MIS $I$, otherwise $I\cup\{v\}$ is an independent set of larger cardinality. If $w$ is in $I$, one can obtain an independent set of the same size by removing $w$ from $I$ and adding~$v$ instead.
Using their branch-and-reduce algorithm, Akiba and Iwata were able to solve a large variety of instances including social networks, web graphs and road networks.
A similar approach that uses a quantum annealer to solve instances once they are small enough was recently presented by Pelofske\etal~\cite{pelofske2019solving}.

Even though Akiba and Iwata~\cite{akiba-tcs-2016} use a sophisticated set of reduction rules, Strash~\cite{strash2016power} showed that many of the more complicated rules are not necessary to compute a MIS in many large complex networks.
Furthermore, the initial reductions applied to compute a kernel often have a bigger impact on performance, compared to further techniques used during the branch-and-reduce approach.
Recently, Stallmann\etal~\cite{stallmann2020graph} supported this idea by showing that networks $G$ with a small normalized average degree ($\mathsf{nad}(G)$) can be efficiently handled by simple reductions.
The $\mathsf{nad}(G)$ of a network $G$ on $n$ vertices is defined as the average degree of $G$ normalized using a factor of $200/n$ if the average is larger than $20$.
Otherwise, if the average degree is at most $20$, $\mathsf{nad}(G)$ is the same as the average degree of $G$.
Additionally, the authors make use of the so-called degree spread $t/b$, where $t$ is the degree at the $95$th percentile and $b$ at the $5$th percentile.
Based on these characteristics, the authors devise thresholds that indicate (1) if reductions should be used at all, (2) if more complex rules provide a significant benefit.

\begin{openproblem}
What are graph characteristics and properties that determine the success of specific reduction rules? 
\end{openproblem}

Recently, Hespe\etal~\cite{hespe2020wegotyoucovered} won the PACE Challenge 2019 vertex cover track by using a portfolio of exact approaches for MIS, MVC and maximum clique.
In particular they use the reduction rules of Akiba and Iwata as an initial preprocessing step.
Afterwards, an initial solution is computed using the state-of-the-art local search algorithm by Andrade\etal~\cite{andrade-2012}.
Finally, they switch between the branch-and-reduce algorithm of Akiba and Iwata~\cite{akiba-tcs-2016} and the clique solver by Li\etal~\cite{li2017minimization} which are applied to either the original graph or the graph resulting from the preprocessing~step.

\subsubsection{Heuristic Approaches}
Reductions are also heavily used in many state-of-the-art heuristic approaches.
Lamm\etal~\cite{DBLP:conf/wea/LammS015,redumis-2017} use the same set of reductions originally used by Akiba and Iwata to develop an evolutionary algorithm that is able to compute high quality solutions for large graphs that are infeasible for branch-and-reduce.
The authors use reductions for both preprocessing (to compute a kernel) and during the algorithm itself.
In particular, they select vertices that are part of many highly fit individuals, \ie~independent sets) in their population.
These vertices are then added to the resulting vertex set, which includes removing them and their neighbors from the graph. 
Afterwards, reduction rules are applied and the evolutionary algorithm is called recursively on the resulting graph.

The idea of excluding a subset of vertices that are likely to be part of a high-quality independent set, is also explored by Gao\etal~\cite{gao2017scaling}.
To select these vertices they perform multiple runs of a state-of-the-art local search (either NuMVC~\cite{numvc-2013} or FastVC~\cite{cai2015balance}).
Vertices that are present in all resulting solutions are then added to the final solution and a new graph consisting of the remaining vertices and their corresponding edges is constructed.
Afterwards, a final run of the local search on this graph is executed and its solution is combined with the previously removed~vertices.

Dahlum\etal~\cite{dahlum2016accelerating} combine both simple exact reduction rules as well as inexact reductions with the ARW local search algorithm~\cite{andrade-2012}.
In particular, they remove cliques of up to size three (exact reductions) and the top 1\% high-degree vertices (inexact reduction).
The reasoning behind their inexact reduction is that high-degree vertices are not likely to be in a large independent set.
Additionally, these vertices pose a significant bottleneck for the local search.
The authors also compare their algorithm against an algorithm that uses the kernelization rules of Akiba and Iwata as a preprocessing step.
A similar preprocessing approach that only uses a subset of reduction rules is also presented by Cai\etal~\cite{cai2017finding}.
In particular, they use the degree-0, degree-1, degree-2 and domination rules.

Chang\etal~\cite{chang2017} also make use of the idea of combining simple reduction rules that can be applied in (near-)linear time with an inexact reduction rule that removes high degree vertices.
For this purpose, they introduce the reducing-peeling framework that switches between the two types of reductions.
Furthermore, they present a set of degree-2 path reductions that are special cases of the folding reduction.
Combining these new rules with the degree-0, degree-1, dominance and an LP-based reduction rule, they propose an efficient preprocessing algorithm that is then combined with the ARW local search.

\begin{openproblem}
Can one derive (near-)linear time special cases of the more complex reductions like the unconfined reduction that are not covered by existing reductions? 
\end{openproblem}

In order to quickly achieve smaller kernels than is possible with simple reduction rules, Hespe\etal~\cite{hespe2019scalable} provided the first shared-memory kernelization based on the rules of Akiba and Iwata.
For this purpose they make use of both graph partitioning and parallel bipartite maximum matchings.
The graph partitioning library KaHIP~\cite{DBLP:journals/corr/SandersS13} is used to compute a partition of that graph which allows  parallel execution of reduction rules that only need to check highly localized subgraphs, where bipartite maximum matchings are used to enable the parallel execution of the LP-based reduction rule.
Furthermore, the authors present two speedup techniques for kernelization: (1) dependency checking that prunes applicability checks for certain reductions and (2) reduction tracking that stops their algorithm once the application of reduction rules only decreases the graph size by an negligible amount.

\begin{openproblem}
Can the techniques used by Hespe\etal~\cite{hespe2019scalable} be extended to a distributed memory setting? How can one efficiently apply reductions in distributed memory?
\end{openproblem}

Alsahafy and Chang~\cite{alsahafy2019computing} recently proposed an algorithm that combines the reducing-peeling framework with the exact clique solver MoMC by Li\etal~\cite{li2017minimization}.
Their algorithms splits reduction rules in two sets: ones that can be updated and applied incrementally (similar to Hespe\etal~\cite{hespe2019scalable}), and ones that can not.
Additionally, they continuously compute and maintain the connected components of the graph, which are then reduced individually.
If a reduced component is small enough, it is then transformed into its complement and solved by MoMC.
To ensure that components continue to get smaller, they use the same inexact reduction rule as Chang\etal~\cite{chang2017} and then continue recursively on the resulting components.
Finally, the authors also present a new exact reduction rule called the pyramid reduction.

Lastly, Lavallee\etal~\cite{DBLP:conf/alenex/LavalleeRSP20} evaluated a structural rounding approach for vertex cover. The main idea is to first edit a graph to a well-structured graph which can be solved more easily, and then apply a ``lifting'' algorithm to the partial solution to recover an approximation on the input network. Lavallee\etal~find that their algorithm can outperform standard 2-approximation algorithms and that simpler lifting strategies are highly competitive with more sophisticated strategies.
\subsubsection{Weighted MIS}
Due to the significant practical results achieved for the unweighted case, there has been an increasing interest in generalizing these techniques for the weighted maximum independent set (WMIS) and weighted minimum vertex cover (WMVC) problems.
For both problems one is given an additional real-valued vertex weighting function $w: V \rightarrow \mathbb{R}^+$. 
In case of the WMIS problem one is then tasked with finding an independent set, such that the sum of the weights of its vertices is maximal among all possible independent sets.
Analogously, for the WMVC one is tasked with finding a vertex cover of minimum weight.

Recently, Li\etal~\cite{li2019numwvc} used a set of four reduction rules during the initial construction phase of a local search algorithm.
In particular, they use weighted reduction rules that are able to remove degree one and degree two vertices.
They then use these reduction rules exhaustively in the beginning of their algorithm to obtain an improved initial solution.
Their local search algorithm called NuMWVC is able to compute high quality solutions on a large variety of instances.
This includes many instances commonly used for the unweighted case, which have been given vertex weights drawn from a uniform distribution.
Since there are not many publicly available weighted instances, this is a common approach that is also used in other works~\cite{cai-dynwvc,gellner2020engineering,lamm-2019,zheng2020efficient}

Wang\etal~\cite{wang2019exact} also make use of reduction rules for vertices with degree at most $2$ as a preprocessing step for a branch-and-bound solver.
Furthermore they evaluate different degree-based heuristics for selecting branching vertices and use pruning based on the best solution found so far.

Lamm\etal~\cite{lamm-2019} proposed the first practically efficient branch-and-reduce algorithm for the WMIS problem that is able to solve a large number of real-world instances.
For this purpose they develop a comprehensive set of practically efficient reduction rules.
This includes both generalization of previous weighted and unweighted reduction rules, as well as two so-called meta reductions which serve as a general framework for the other rules.
They use these rules to build a branch-and-reduce algorithm that uses many of the approaches that worked well in the unweighted case.
In particular, they use local searches to compute initial solution which can be used for pruning, treat connected components individually and make use of dependency checking.
Finally, they show that their reduction rules can be used to improve the performance of other state-of-the-art algorithms

Zheng\etal~\cite{zheng2020efficient} propose an exact and heuristic approach that both make use of reduction rules for vertices of degree at most $2$.
Their exact approach is a branch-and-reduce algorithm that applies these reduction rules recursively. 
However, the authors do not provide any details on the bounds or branching strategies used during the algorithm.
Their heuristic approach is inspired by the reducing-peeling framework of Chang\etal~\cite{chang2017}.
Thus, it exhaustively applies their reduction rules and subsequently removes high-degree vertices to extend the space of possible reductions.

Gellner\etal~\cite{alex2020boosting} proposed new practically efficient variants of the struction rule by Ebenegger\etal~\cite{ebenegger1984pseudo}.
The struction is a reduction that is able to be applied to arbitrary vertices in a graph, but comes at the cost of potentially increasing the overall number of vertices.
Thus the authors propose three new variants of the struction that aim to limit the number of newly created vertices.
Furthermore, they derive practically efficient special cases of their reduction rules and use them as a preprocessing step in the branch-and-reduce solver of Lamm\etal~\cite{lamm-2019}.
The algorithm is able to produce the smallest-known reduced graphs, solves more instances than previous exact approaches and has a running time that is comparable to~heuristic~algorithms.

\begin{openproblem}
Can other problems also benefit from reductions that may temporarily increase the graph size? If so, how much of an increase should be allowed to remain practical?
\end{openproblem}

\subsection{Finding and Enumerating Maximum Cliques}
Given an undirected graph $G=(V,E)$, the goal of the \emph{maximum clique} (MC) problem is to compute a set of vertices $C \subseteq V$ such that (1) all vertices in $C$ are adjacent to one another, (2) the set $C$ has maximum cardinality among all such sets.
As mentioned in the previous section, MC solvers are often used in the context of independent sets.
This is due to the fact that a clique of $G$ is an independent set in the complement graph $\bar{G} = (V, \bar{E})$ with $\bar{E} = \{\{u,v\} \mid u,v \in V \land \{u,v\} \not\in E\}$.
Likewise, finding maximum cliques is an NP-hard optimization problem~\cite{garey1990guide}.
Furthermore, unless P$=$NP, there is no polynomial time approximation for MC that can provide an $\Oh{n^{1-\epsilon}}$ guarantee for any constant $\epsilon >0$~\cite{zuckerman2006linear}.
Finally, MC is $W[1]$-hard~\cite{downey1999parameterized} parameterized by solution size $k$, making it unlikely that the problem is fixed-parameter tractable in $k$.
However, it is fixed-parameter tractable under different parameterizations, \eg~when parameterized with the degeneracy of the graph~\cite{eppstein2013listing}.
All previous observation also hold for the \emph{maximum clique enumeration} (MCE) problem of enumerating all maximum cliques in a graph.

Eblen\etal~\cite{eblen2012maximum} presents a maximum clique solver (MCF) that adapts some of the reduction rules that have already been shown to work well for MVC and MIS. 
In particular, their algorithm begins by greedily computing a large clique $C$ which is then used as a lower bound in order to remove vertices of degree less than \mbox{$|C|-1$}~\cite{pardalos1999maximum}.
Next, they use an adaptation of the degree-0 reduction rule previously used in MVC algorithms, as well as a rule based on heuristic colorings~\cite{tomita2007efficient} to remove additional vertices. 
The authors also investigated the use of other reduction rules including an adaptation of the degree-1 reduction rule used in MVC algorithms.
Finally, they compared applying reduction rules as a preprocessing method for a branch-and-bound solver against running them in a branch-and-reduce solver.
Their experiments indicate that the branch-and-reduce approach performs better on real-world genome data.

Eblen\etal~\cite{eblen2012maximum} then used the previous MCF solver to develop several approaches for the maximum clique enumeration (MCE) problem based on the algorithm by Bron and Kerbosch~\cite{bron1973algorithm}.
In particular, they develop two reduction rules based on MCF:
First, they propose a reduction rule that uses MCF to compute a maximum clique cover and removes vertices not adjacent to this cover.
Second, they propose a second data-driven preprocessing rule that computes so-called essential vertices, \ie~vertices that are present in every maximum clique. 
Vertices that are not adjacent to these vertices are subsequently removed from the graphs.
Their experiments indicate that this rule works particularly well on large transcriptomic graphs, that often have a small set of essential vertices.
However, it performance degrades for networks that do not have a small set of essential vertices, \eg~for uniform random graphs.

\begin{openproblem}
Can one give similar data-driven reduction rules for other types of networks, \eg~social networks or road networks? 
\end{openproblem}

Verma\etal~\cite{verma2015solving} propose another type of reduction rule based on so-called $k$-communities.
For this purpose, a $k$-community subgraph is defined as a subgraph $G'=(V',E')$ where each edge $\{u,v\} \in E'$ connects vertices that have at least $k$ common neighbors in $G'$.
Subsequently, a subset of vertices $V' \subseteq V$ is called a $k$-community if there is a $k$-community subgraph with vertex set $V'$ in $G$.
Note, that a clique of size $k$ is a $(k-t)$-community for any $t\in\{2,\dots,k\}$.
They then derive a reduction rule which computes a lower bound on the clique size based on maximum $(k-2)$-communities and prune vertices with a smaller degree.
They then combine this reduction rule with the $k$-core based approach of Pardalos and Resende~\cite{pardalos1999maximum} and show that the resulting algorithm works well for handling large low-density graphs.

Chang~\cite{chang2019efficient,chang2020efficient} notes that even though a lot of real-world networks are usually sparse, MC has been more extensively studied for dense instances.
Thus, the authors propose a branch-and-reduce algorithm that leverages the existing work on MC for dense instances by transforming an instance of MC over a sparse graph to instances of $k$-clique finding (KCF) over dense subgraphs.
For this purpose, the authors iteratively compute small and dense subgraphs (so-called ego networks) that are then handled by a KCF solver.
In order to reduce the size of the subgraphs that are handled by this solver, their algorithm uses a combination of well-known upper bounds and lightweight reduction rules.
In particular, they use five reduction rules for KCF, most of which are targeted toward removing vertices of high degree.
The authors also present a heuristic algorithm for MC, as well as two stage approach for MCE that makes use of their exact algorithm to compute the size of the largest clique. 
Furthermore, they show that the reduction rules used for MC can also be adapted for MCE.

\subsubsection{Weighted MC}
Recently, Cai and Lin~\cite{cai2016fast} proposed the first (and only) practical algorithm for the \emph{(vertex-)weighted maximum clique} (WMC) problem that uses reduction rules. 
The WMC problem is a generalization of MC where one is given an additional real-valued vertex weighting function $w: V \rightarrow \mathbb{R}^+$.
Subsequently, one is tasked with finding a clique, such that the sum of the weights of its vertices is maximal among all possible cliques.
In order to solve WMC on large sparse graphs, Cai and Lin~\cite{cai2016fast} interleave clique construction with reduction rules.
To be more specific, they gradually add ``beneficial'' vertices to a clique using an approximation of the benefit of a vertex.
This is done by computing the mean of a cost-efficient upper and lower bound for each vertex and then selecting vertices using a dynamic best from multiple selection~\cite{cai2015balance}.
Finally, if a new best clique is found, the graph is reduced using two reduction rules.
Both rules make use of the fact that one is able to remove vertices where an upper bound on any maximum clique containing this vertex is smaller than the weight of the current best clique.
For their rules, the authors then propose two different upper bounds that make use of the neighborhood of a vertex.

\subsubsection{$k$-plexes}
$K$-plexes are a generalization of cliques where each vertex is allowed to have several missing connections, \ie~not every vertex has to be connected to all other vertices in the $k$-plex~\cite{seidman1978graph}.
In particular, a $k$-plex is a subset $S \subseteq V$ such that the degree of every vertex in the induced subgraph $G[S]$ is at least $|S|-k$.
Furthermore, $|S|$ is called the size of the $k$-plex and the \emph{maximum $k$-plex problem} (MK) is that of finding a $k$-plex of maximum size.

Gao\etal~\cite{gao2018exact} present multiple theoretical properties that allow the removal of vertices based on a lower bound on the maximum $k$-plex size. 
Based on these properties they propose four reduction procedures which are then used in a branch-and-reduce algorithm. 
In particular, they then use an extension of the algorithm by Jiang\etal~\cite{jiang2017exact} to compute an initial lower bound and then use this bound to exhaustively apply their linear-time vertex reduction and the more costly subgraph reduction rules for preprocessing.
Afterwards they use different sets of reduction rules depending on the type of branch (selecting or discarding a vertex).
The authors also present a type of targeted branching that aims to select vertices which lead to a larger reduction in size.
The resulting algorithm is able to solve multiple previously infeasible real-world instances and is considerably faster than previous state-of-the-art solvers (\eg~\cite{xiao2017fast}).

\begin{openproblem}
Can targeted branching be used for other problems? For example, the most commonly used branching strategy for independent sets is degree-based and does not take any reduction rules into account. 
\end{openproblem}

Conte\etal~\cite{conte2017fast} investigated reduction rules for the problem of enumerating all maximum $k$-plexes.
For this purpose, they introduce the concepts of coreness and cliqueness.
Coreness states that vertices of a $k$ plex of size at least $m$ must have a degree not smaller than $m-k$.
Thus, vertices with a smaller degree can iteratively be removed, resulting in the computation of $(m-k)$-cores.
Cliqueness states that every vertex of a $k$-plex of size at least $m$ is part of a clique not smaller than $\lceil m/k \rceil$.
Therefore, vertices with a degree less than $\lceil m/k \rceil$ can be removed from the graph.
Furthermore, if one knows the size of the maximum clique $\omega$ the search space for the size of the maximum $k$-plex can be limited to $[\omega, \omega \cdot k]$.
Based on these concepts the authors then present an algorithm that begins by computing the size of the maximum clique. 
Afterwards a lower bound for the size of the maximum $k$-plex $p \in [\omega, \omega \cdot k]$ is guessed.
If this guess turned out to be wrong (\ie~all $k$-plexes found are smaller than~$p$), the interval bounds are updated and a new lower bound is guessed.
Otherwise, all $k$-plexes with maximum size are returned.
Their algorithm is able to reduce a large set of instances by up to $99\%$ and achieves running times that are multiple orders of magnitude faster than previous approaches~\cite{berlowitz2015efficient}.

\subsection{Maximum Cuts}
The \emph{max-cut} problem originates from important applications in physics and operations research~\cite{BarahonaGJR1988}; therefore, it has long been the subject of engineering more and more sophisticated algorithms which solve large-scale instances arising in practice.
In particular, {max-cut} is one of the few problems where engineers and practitioners alike are interested in finding optimal solutions (rather than just approximate ones).
Formally, the {max-cut} problem takes as input an edge-weighted graph $G$ and seeks a bipartition of the vertex set $V$ of $G$ into two disjoint parts, $V_1$ and $V_2$, which maximizes the weight of the edges which \emph{cross} the bipartition, that is, edges whose one endpoint is in $V_1$ and the other endpoint is in $V_2$.
The state of the art for {Max-Cut} though is that even after much effort, optimal solutions are still unknown for several benchmark instances.
Those reasons are the key motivations for engineering effective, and efficient, kernelization rules.
The objective is to reduce the given graph~$G$ to a new instance~$G'$ of smaller size, such that a maximum cut in $G$ can be recovered efficiently from any maximum cut in~$G'$.
To the best of our knowledge, preprocessing rules with theoretical guarantees have been studied so far mainly for the unit-weight {max-cut}.
That special case of {max-cut}, where all edges have the same (unit) weight, is still NP-hard.
The goal is thus to find a bipartition $(V_1,V_2)$ which maximizes size of the cut, which is the number of edges with one endpoint in $V_1$ and the other endpoint in $V_2$.
To measure the effectiveness of preprocessing rules for unit-weight {max-cut}, one introduces an integer parameter~$k$.
This parameter measures the difference between the size of the maximum cut, and the value $m/2 - (n-1)/4$, which is the well-known lower bound on the size of the maximum cut in any $m$-edge $n$-vertex graph, due to Edwards and Erd{\H{o}}s~\cite{Edwards1973a,Edwards1973b}.
There is a set of preprocessing rules, devised by Etscheid and Mnich~\cite{EtscheidM2018} which compresses any $m$-edge $n$-vertex graph $G$ in linear time to a graph $G'$ on just $\Oh{k}$ vertices, while allowing to recover the maximum cut of $G$.
This set of rules strengthened earlier work by Crowston\etal~\cite{CrowstonJM2015}, and is moreover asymptotically best possible.
To understand the practical relevance of those rules, Ferizovic\etal~\cite{FerizovicHLMSS2020} expanded and engineered them.
They demonstrated their significant impact on benchmark data sets, including synthetic instances, and data sets from the VLSI and image segmentation application domains.
Their experiments revealed that current state-of-the-art solvers can be sped up by up to multiple orders of magnitude when combined with their data reduction rules.
On social and biological networks in particular, the preprocessing enabled them to solve four instances that were previously unsolved in a ten-hour time limit with state-of-the-art solvers; three of these instances are now solved in less than two seconds.
It is possible to expand the work on preprocessing for unit-weight {max-cut} to instances with all positive weights.
However, designing practically efficient preprocessing rules for the general {max-cut} problem, which also provides theoretical guarantees on the kernel size, remains a challenge.
Recent work in this direction was done by Lange\etal~\cite{LangeAS2019}, who designed reduction rules for general {max-cut}.
They showed the efficacy of their rules on instances from computer vision, biomedical image analysis and statistical physics, and for those instances managed to obtain substantial size reductions.

\begin{openproblem}
Is it possible to engineer efficient reduction techniques for max-cut with general edge weights?
\end{openproblem}

\subsection{Treewidth and Treedepth}
\newcommand{\tw}[0]{\ensuremath \mathsf{tw}(G)}
\newcommand{\td}[0]{\ensuremath \mathsf{td}(G)}

Many NP-hard graph problems can be efficiently solved when the input graph is a tree. A tree decomposition maps vertices of a graph to vertices in a tree, which allows techniques for trees, especially dynamic programming, to be adapted to arbitrary graphs. However, the quality of the tree decomposition impacts the efficiency of such algorithms. \emph{Treewidth}~\cite{robertson1986graph} is one measure of this quality, which has been extensively studied in parameterized algorithms literature, which we now describe.

Formally, a \emph{tree decomposition} of a graph $G=(V,E)$ is a family of subsets $\mathcal{X}\subseteq 2^{V}\setminus\{\varnothing\}$ of $V$ called bags, together with a tree $T = (\mathcal{X}, F)$, such that
\begin{itemize}
\item $V=\cup_{X\in\mathcal{X}} X$,
\item for all $\{u,v\}\in E$ there exists a bag $X\in\mathcal{X}$ such that $u,v\in X$, and
\item for all $v\in V$, the bags $\mathcal{X}_v = \{X\in \mathcal{X} \mid v\in X\}$ containing $v$ induce a connected subgraph $T[\mathcal{X}_v]$ (which is necessarily a subtree of $T$). 
\end{itemize}

The \emph{width} of a tree decomposition of $G$ is one less than the cardinality of its largest bag, that is, $\max_{X\in\mathcal{X}}\{|X|\} - 1$. The treewidth of $G$, denoted $\tw$, is the minimum width over all tree decompositions of $G$.

Unsurprisingly, computing $\tw$ is NP-hard and deciding if $\tw \leq k$ for some positive integer $k$ is NP-hard. The treewidth problem is a canonical problem with many theoretical and practical results in the literature. In particular, it is fixed-parameter tractable with running time $2^{\Oh{k^3}}n$~\cite{bodlaender1996linear}, implying it has a kernel exponential in $k^3$~\cite{cai1997advice}. The problem does not have a kernel size subexponential in $k$ unless \mbox{NP $\subseteq$ coNP/poly}~\cite{bodlaender2009on}. Hence, most work focuses on constructing tree decompositions of small width, either approximately~\cite{bodlaender2016a}, or exactly using methods such as positive-instance driven dynamic programming~\cite{tamaki2017positive}. Both the first and second PACE Challenges had a treewidth track~\cite{dell2017pace} 
However, polynomial kernels exist for other parameters. Bodlaender\etal~\cite{bodlaender2013preprocessing} give polynomial kernels of size $\Oh{\mathsf{fvs}(G)^4}$ and $\Oh{\mathsf{vc}(G)^3}$, where $\mathsf{fvs}(G)$ is the size of a minimum feedback vertex set and $\mathsf{vc}(G)$ the size of a minimum vertex cover of $G$, respectively. Their work is inspired by data reduction rules that are known to work well in practice (discussed below), and also includes new rules based on the notion of `clique seeing' paths. Jansen~\cite{jansen2015on} improved the latter kernel to size $\Oh{\mathsf{vc}(G)^2}$ by introducing a new reduction rule to efficiently find independent sets whose elimination has a predictable effect on the treewidth. To the best of our knowledge, no experiments have been done with clique seeing paths or Jansen's reduction.

\begin{openproblem}
Is the rule of Jansen~\cite{jansen2015on} effective in practice?
\end{openproblem}

Much work has been done in making practical data reductions for the treewidth problem. In early work, Arnborg and Proskurowski~\cite{arnborg1986characterization} introduced reduction rules for recognizing and characterizing partial 3-trees. Bodlaender\etal~\cite{bodlaender2005preprocessing} categorized these reductions into six types (islet, twig, series, triangle, buddy, and cube) and extended these rules, showing them to be highly effective at reducing graph size in practice~\cite{bodlaender2005preprocessing}. Of note here are two variations of well-known reductions from other problems: simplicial vertices and twins of degree 3. They further give a reduction for \emph{almost} simplicial vertices (vertices with all but one neighbor inducing a clique).  On graphs with up to 3\,032 vertices, the reductions quickly remove 77\% of vertices on average, whereas the simplicial vertex reduction alone remove 51\% of vertices on average. The worst performing instances had 30\% of its vertices removed.
Den van Eijkhof\etal~\cite{eijkhof2007safe} generalized many of these reduction rules. They not only introduce new weighted variants, but generalize most previous reductions with a `contraction' reduction rule, and further introduce a reduction for twins of higher degree.

Later, Bodlaender\etal~\cite{bodlaender2006safe} introduced the concept of a safe separator, which decomposes the graph into subgraphs that can be solved independently. It was already known that clique separators were safe~\cite{olesen2002maximal}; however, they generalize the concept, and introduce other easy-to-find separators. They further show that previous reduction rules are subsumed by their safe separator technique. In experiments, their reductions decomposed 33 out of 40 instances. When first run as a preprocessing step, their technique speeds up an existing triangulation heuristic, sometimes by multiple orders of magnitude. However, it only gives modest speedups over preprocessing using existing reductions.

\begin{openproblem}
How effective are existing treewidth reductions on large sparse graphs (e.g., with millions of vertices) in practice?
\end{openproblem}

\begin{openproblem}
Can heuristic methods be used to efficiently find safe separators in practice?
\end{openproblem}

A related concept exists for decompositions into rooted trees. A \emph{treedepth decomposition} of a graph $G = (V,E)$ is a rooted forest $F$, together with an injective mapping $\phi:V(G) \rightarrow V(F)$ such that, for each edge $(u,v)\in E$, one of $\phi(v)$ or $\phi(u)$ is an ancestor of the other. The treedepth of~$G$, denoted by $\td$, is the minimum height of any treedepth decomposition of $G$. The treedepth problem, deciding if $\td \leq k$ for some positive integer $k$, is NP-hard~\cite{pothen1988the}.

Given their similarity, it is perhaps unsurprising that many similar results exist for the treewidth and treedepth problems. Reidl\etal~\cite{reidl2014a} give a fixed-parameter tractable algorithm for treedepth $k$, with running time $2^{\Oh{k^2}}n$, implying the existence of a kernel of size exponential in $k^2$, and no subexponential kernel exists unless NP $\subseteq$ coNP / poly~\cite{bodlaender2009on}. However, when parameterized on the vertex cover number $\mathsf{vc}(G)$, the problem has a kernel of size $\Oh{\mathsf{vc}(G)^3}$~\cite{kobayashi2017treedepth}, which is achieved through two simple reduction rules that also apply to treewidth: removing simplicial vertices and adding edges between vertices with at least $k$ common neighbors.

However, as far as we are aware, there are significantly fewer experimental works with data reduction rules for treedepth. The 5th PACE Challenge in 2020 was dedicated to exact and heuristic solutions for treedepth. The winning solver by Trimble~\cite{trimble2020algorithm} did not employ any data reduction rules (instead, using symmetry breaking together with a variety of lower bounding techniques); however, the second place solver by Korhonen~\cite{korhonen2020sms} applies the simplicial vertex rule by Kobayashi and Tamaki~\cite{kobayashi2017treedepth} and a generalization of their common neighbor rule. Korhonen further introduces a new reduction rule based on the Sch\"affer's linear-time algorithm~\cite{schaffer1989optimal} for computing the treedepth of trees. This rule replaces a tree subgraph $G[T]$ having $|N(V\setminus T)|=1$ with a subgraph of size $\mathsf{td}(G[T]^2)$. As far as we know there are no published results on the efficacy of these reduction rules. Of further interest is that this algorithm uses minimal separator enumeration. We conclude with the following open problems.

\begin{openproblem}
How effective are the reductions of Kobayashi and Tamaki~\cite{kobayashi2017treedepth} and Korhonen~\cite{korhonen2020sms} in practice?
\end{openproblem}

\begin{openproblem}
Does the notion of a safe separator extend to the treedepth problem?
\end{openproblem}

\subsection{Hitting Set}
    
Given a set $S$ along with a collection $C$ of its subsets, the \emph{hitting set} problem asks for a subset of $S$, of minimum cardinality, that has a non-empty intersection with each and every member of $C$. 
Hitting set is the dual of \emph{set cover}, which seeks a minimum-cardinality subset of $C$ whose union is $S$. 
If the elements of $S$ and $C$ are treated as red and blue vertices, respectively, of a bipartite graph, the equivalent graph theoretic problem is known as \emph{red-blue dominating set} (RBDS).

Hitting set is NP-hard, and $W[2]$-hard when parameterized by the solution size~\cite{downey1999parameterized}. It becomes fixed-parameter tractable when each member of~$C$ is of size bounded by a constant $d$. In this case the problem is often referred to as~$d$-Hitting Set and it corresponds to RBDS restricted to (red-blue) graphs where each red vertex has at most $d$ neighbors. The problem is also known to be fixed-parameter tractable when parameterized by $|C|$, but this particular parameter is expected to be large in practice.
The most popular reduction procedures for Hitting Set are due to Weihe~\cite{Weihe}. They are simply based on removing any possible redundant elements from $S$ and $C$. In this context, an element of $S$ is redundant if all members of~$C$ that contain it contain another element; while a member of $C$ is redundant if it is a superset of another member of $C$. The application of these two rules alone proved to be highly effective on large public transportation networks resulting in a huge reduction in size as pointed out recently in \cite{BlasiusF0S19}.

More sophisticated reduction algorithms appeared in the context of kernelization for \emph{$d$-Hitting Set}~\cite{Abu-Khzam10,Moser2010, NiedermeierR03}. 
The kernelization approach of
\cite{Abu-Khzam10} was adopted in \cite{Mellor2010} and proved to be effective
in the context of multiple drug selection for cancer therapy.
Moreover, linear-time algorithms that can obtain a kernel of size $\Oh{k^d}$ have been presented in \cite{Bevern2014} and \cite{FK2015}.
Practical implementations of these algorithms have been addressed recently in \cite{Bevern2020} where they were shown to be more efficient than the reduction procedure of \cite{Weihe} for small $d$ (up to 5), but can result in more effective data reduction when combined with the reduction rules of \cite{Weihe}.

\subsection{Steiner Trees}
Given an undirected graph with non-negative edge weights as well as a subset of the vertices (terminals), the \emph{Steiner tree} problem is to find the lightest tree spanning the terminals. There has been a wide range of implementations tackling the Steiner tree problem. Data reductions have long been used for the problem, see, \eg~Polzin~\cite{DBLP:phd/de/Polzin2004} or Daneshmand~\cite{DBLP:phd/de/Vahdati-Daneshmand2004}. Daneshmand~\cite{DBLP:phd/de/Vahdati-Daneshmand2004} in particular shows already in 2004 that many Steiner tree problem instances can be solved by reduction- and heuristic-based approaches. 

Recently there have been two implementation challenges, the 11th DIMACS Challenge in 2014 and the 3rd PACE Challenge~\cite{DBLP:conf/iwpec/BonnetS18} in 2018.  Here, we focus on the most successful implementations of the recent PACE Challenge and the approaches that have been published afterwards. The PACE challenge had three tracks overall -- two exact tracks with one focusing on algorithm for problems with few terminals and one focusing on problems with low treewidth, as well as one heuristic track.

The implementation of Iwata and Shigemura~\cite{DBLP:conf/aaai/IwataS19} won the track with problems that have few terminals. Their algorithm is based on the dynamic programming formulation of Erickson-Monma-Veinott~\cite{DBLP:journals/mor/EricksonMV87} which has a  theoretical running time of $\Oh{3^tn + 2^t(n \log n + m)}$ with $t$ being the number of terminals. Iwata and Shigemura use a novel separator-based pruning technique to speed up their implementation (while keeping the worst-case bound of Erickson-Monma-Veinott). This technique allows them to prune a large amount of entries in the dynamic programming table. 

The track with problems that have low treewidth has been won by SCIP-Jack~\cite{DBLP:conf/or/RehfeldtK17,DBLP:journals/networks/RehfeldtKM19} due to Koch and Rehfeldt. This approach is based on the branch-and-cut principle and has already been very successful during the DIMACS Challenge on the problem. For the PACE Challenge, the authors use data reductions that typically reduce the number of edges in the problems by more than 90\%. Many instances can already be completely solved by presolving. Moreover, on the remaining instances that can not be presolved, the authors use heuristics to find strong upper and lower bounds quickly. The authors find that in more than 90\% the heuristic already finds the optimum solution on the instances that have not been presolved. Lastly, the branch-and-cut procedure is used to compute lower bounds and prove optimality. Later, the approach has been improved \cite{DBLP:conf/cpaior/ShinanoRK19} to run in distributed memory and thus, by using up to 43\,000 cores, managed to solve additional previously unsolved instances or improved on the previously best known solution.
\begin{openproblem}
Are there new reductions that have not yet been tried in practice that could help to solve more instances to optimality in practice?
\end{openproblem}
\begin{openproblem}
Can existing reductions for the standard Steiner tree problem be transferred to the more general multi-level Steiner tree problem?
\end{openproblem}
\subsection{Minimum Fill-In}
The \emph{minimum fill-in problem} is a critical problem that accelerates Gaussian elimination when solving sparse linear systems~\cite{rose1970triangulated}. Given an matrix $A$ representing the sparse linear system $Ax = b$, the goal is to find a permutation matrix $P$ that minimizes the number of non-zeros introduced when factorizing $A = PAP^{T}$. Equivalently, treating $A$ as the adjacency matrix of a graph $G=(V,E)$, we wish to minimize the number of edges introduced in an \emph{elimination ordering}, defined as follows. An \emph{elimination step} removes a vertex $v\in V$ and its incident edges, and adds edges between non-adjacent vertices in $N_G(v)$, producing an elimination graph $G_v$. An \emph{elimination ordering} of $G$ is a permutation $v_1v_2..v_n$ of all the vertices in $G$, and the \emph{fill-in} of the ordering is the number of edges introduced by eliminating vertices $v_1,v_2,\ldots,v_n$ in this order. The minimum fill-in is the smallest fill-in given by any elimination ordering. We are often interested in not just computing the minimum fill-in, but an elimination ordering that has minimum fill-in. 

Not only is minimum fill-in NP-hard to compute~\cite{yannakakis1981computing}, no polynomial-time approximation scheme exists for the problem unless P=NP~\cite{cao2020minimum}. However, the problem
is fixed-parameter tractable~\cite{kaplan1994tractability}, when the input parameter
$k$ is the minimum fill-in. The fastest known fixed-parameter algorithm for the problem is due to Fomin and Villanger~\cite{fomin2013sub}, with running time $2^{\Oh{\sqrt{k} \log k}} + \Oh{k^2nm}$, where the additive $\Oh{k^2nm}$ term is the time to compute a kernel of
$\Oh{k^3}$ vertices via the algorithm of Kaplan\etal~\cite{kaplan1999tract}. 
Note that this algorithm is subexponential in the minimum fill-in $k$ and, moreover, is nearly optimal: Cao and Sandeep~\cite{cao2020minimum} showed that no algorithm with running time $2^{\Oh{k^{1/2-\delta}}}\cdot n^{\Oh{1}}$ exists for any positive constant $\delta$, assuming the exponential time hypothesis holds. The smallest known kernel for the problem is due to Natanzon\etal~\cite{natanzon2000a} has $2k^2 +4k$ vertices. The reductions all have the same flavor, and are derived for the equivalent problem of \emph{chordal completion}: finding the minimum number of edges to add to the graph so that it is chordal. Kernelization is done by partitioning the vertices into two sets $A$ and $B$ where $B$ induces a chordal graph and $A$ contains vertices from every chordless cycle in $G$. The set $A$ is formed by repeatedly finding chordless cycles in $G[B]$ via the MCS algorithm~\cite{tarjan1984simple,tarjan1985addendum} and moving a subset of their vertices to $A$ until $G[B]$ is chordal. Then so-called \emph{essential edges} are added to the chordless cycles induced by $A$, which is the kernel.

In practice, the minimum fill-in problem is extremely hard
to solve exactly. Indeed, in the second PACE Challenge in 2017, the winning
solver for the minimum fill-in problem only solved
54 out of 100 instances~\cite{dell2017pace}, when each instance is given a 30-minute time limit. The top three submissions all used kernelization~\cite{kaplan1999tract} together with dynamic programming over potential maximal cliques~\cite{bouchitte2001treewidth,tamaki2017positive}. The first place submission by Kobayashi and Tamaki used generalized variants of the data reduction rules of Bodlaender\etal~\cite{bodlaender2011faster}, and the third place submission performed preprocessing adapted from the safe separator technique for treewidth~\cite{bodlaender2006safe} in addition to kernelization~\cite{kaplan1999tract}. 

However, heuristic techniques, including nested dissection~\cite{george1973nested} and minimum-degree ordering~\cite{tinney1967direct}, work quite well in practice for real-world (typically sparse) graphs. Early researchers noted that indistinguishable vertices may be eliminated together, and therefore may be collapsed into a representative vertex while ordering~\cite{ashcraft1995compressed,duff1996exploiting}. This reduction speeds up the minimum degree algorithm by more than a factor two in experiments~\cite{george1989evolution}. Ost~\etal~\cite{ost2020engineering} recently introduced new data reduction rules based on twins, simplicial vertices, and path compression, and experiments show that they are highly effective in practice when applied before running nested dissection. For road networks, when used as a preprocessing step with other inexact reductions, their techniques give speedups of between 1.79 and 6.37 over nested dissection while simultaneously reducing the fill-in. On social networks, their reductions yield speedups of between 1.72 and 3.92 on 19 out of 21 social networks tested, and the fill-in was reduced on all but one instance.

\begin{openproblem}
How effective are the reductions by Ost~\etal~\cite{ost2020engineering} when combined with other reductions~\cite{natanzon2000a}?
\end{openproblem}

\begin{openproblem}
Is branch-and-reduce feasible for the minimum fill-in problem?
\end{openproblem}

\subsection{Vertex Coloring}
Given an unweighted, undirected simple graph $G=(V,E)$, the \emph{$q$-coloring} problem asks if there exists an assignment of at most $q$ colors to all vertices in $V$ such that no two adjacent vertices have the same color (\ie~a \emph{proper coloring}). The problem of finding the minimum number $\chi(G)$ of colors for which a proper coloring of $G$ exists is known as the \emph{chromatic number} problem.

These problems have received considerable attention by the parameterized algorithms community; however, somewhat surprisingly, there is a wide divide between theory and practice. In theory, a kernel parameterized on only the number of colors is unlikely: since graph coloring is NP-hard for $q=3$ colors~\cite{garey1974}, this would give a constant-sized kernel, implying P$=$NP. Therefore, research has focused on~other~parameters.

When considering the treewidth $\tw$ of the graph $G$, if $G$ is given together with a tree decomposition of width $k\geq \tw$, dynamic programming over the tree decomposition gives an algorithm solving $q$-coloring in time $q^kk^{\Oh{1}}n$~\cite[Theorem 7.9]{cygan2015parameterized}. Assuming the Strong Exponential Time Hypothesis (SETH) no algorithm of running time $\Oh{(q-\varepsilon)^{\tw}}$ exists~\cite{lokshtanov2018known} for any $\varepsilon > 0$. Using the same technique, the chromatic number can be computed in time $k^{\Oh{k}}n$~\cite[Theorem 7.10]{cygan2015parameterized}. Since these algorithms are fixed-parameter algorithms, the result due to Cai\etal~\cite{cai1997advice} implies kernels of size $q^kk^{\Oh{1}}$ and $k^{\Oh{k}}$ exist for $q$-coloring and chromatic number, respectively. Treewidth is often small for sparse graphs in practice; however, as far as know, these techniques have not been tried in practice, leading to the following open problem.

\begin{openproblem}
How effective is dynamic programming over a tree decomposition for $q$-coloring (or chromatic number) on sparse graphs in practice?
\end{openproblem}

Another parameter of interest is size of a minimum vertex cover. Recently, Jansen and Pieterse~\cite{jansen2019optimal} gave a kernel parameterized on the number $q\geq 3$ of colors and the size $k$ of a minimum vertex cover, having size $\Oh{k^{q-1}\log k}$ bits, which is optimal up to a factor of $k^{\Oh{1}}$~\cite{jaffke2017fine}. Their result also applies for a tighter parameter, when~$k$ is the size of the twin cover.
Their technique uses constraint satisfaction with low-degree polynomials.  However, in practice sparse graphs often have a minimum vertex cover size that is linear in the number of vertices. Thus, to be useful in practice, the actual kernel would need to have significantly smaller size. However, to date no one has tested their method in practice, leading to our next open problem for~$q$-coloring.

\begin{openproblem}
How effective are the reductions of Jansen and Pieterse~\cite{jansen2019optimal} in practice?
\end{openproblem}

Those data reductions that have been implemented in practice are simple and without theoretical guarantees on the kernel size; however, they are also very effective on large sparse graphs.
In particular, in experiments for a branch-and-cut algorithm, Mend\'{e}z-D\'{i}az and Zabala~\cite{diaz-2006} first preprocess input graph by computing a large clique $K$ of $k$ vertices, which is a lower bound on the chromatic number. They then iteratively remove each vertex $v$ of degree at most $k-1$ (resulting in a $k$-core), which is possible since $\chi(G) = \chi(G - \{v\})$. They also give a variant of the domination reduction, arguing that any vertex $v\in V\setminus N[K]$ dominated by any $w\in K$ (\ie~$N(v) \subset N(w)$) can be removed without changing the chromatic number. In experiments on 63 graphs of up to 5\,231 vertices from the second DIMACS implementation challenge\footnote{\scriptsize\url{http://archive.dimacs.rutgers.edu/pub/challenge/graph/benchmarks/color/}}, their data reductions reduced all graphs between 1--93\%, working best on sparse instances. 36 of the 63 instances were reduced by at least 25\%, and 21 instances were reduced by at least 50\%. The largest percentage reduction was 93\% for the \texttt{homer} instance, reducing from 561 to 38 vertices.

Verma\etal~\cite{verma2015solving} extend this technique for decreasing values of~$k$. They first compute lower and upper bounds for the chromatic number, and then iteratively apply the $k$-core reduction to heuristically color graphs for decreasing values of~$k$. Their key contribution is beginning with an exact coloring of the $k$-core, which gives a better bound than an initial clique. With this technique they are able to exactly find the chromatic number for very large sparse graphs with up to millions of vertices, with running time varying from seconds to hours. In total they solve 33 of 53 instances from SNAP\footnote{\scriptsize\url{http://snap.stanford.edu/data}} and the tenth DIMACS implementation challenge\footnote{\scriptsize\url{http://www.cc.gatech.edu/dimacs10/}}. Lin\etal~\cite{ijcai2017-73} extended the low degree reduction to remove entire independent sets of vertices with low degree, which in some cases is orders of magnitude faster than the algorithm of Verma\etal~\cite{verma2015solving}. However, they are not able to solve any additional instances.

We finally note that a crown reduction exists for the \emph{dual coloring} problem, which asks if the graph has an $(n-k)$-coloring~\cite{fomin2019kernelization}. Crown reductions are particularly effective in practice for other problems, specifically the vertex cover problem. In theory, for dual coloring, the crown reduction produces a kernel of size at most~$3k-3$~\cite[Theorem 4.9]{fomin2019kernelization}. As far as we are aware, no one has performed experiments with this reduction, leading to the final open problem for graph coloring.

\begin{openproblem}
How effective is the crown reduction~\cite[Theorem 4.9]{fomin2019kernelization} for graph coloring in practice?
\end{openproblem}

\subsection{Cluster Editing}
The \emph{cluster editing} problem is as follows: given a graph $G=(V,E)$, transform it into a vertex-disjoint union of cliques by inserting and deleting a minimum number of edges,\ie~by making a minimum number of editions in the graph. The problem is also known as correlation clustering and has many applications, especially in computational biology~\cite{Bocker13}.
The parameterized complexity of the cluster editing problem using the number of edits $k$ as a parameter is well-studied. The currently best known algorithm in theory is due to B\"ocker~\cite{Bocker12} 
and has running time $\Oh{1.62^k+n+m}$, where $m$ is the number of edges. 

There has been a wide range of methods applying fixed-parameter techniques in the area.
Dehne\etal~\cite{10.1007/11847250_2} presented the first practical implementation of a fixed-parameter based method for cluster editing. Their algorithm is an exact algorithm which implements the kernelization routines of \cite{DBLP:conf/ciac/GrammGHN03} and adds ideas to bound the search space for the parameter $k$ via linear programming. Gramm\etal~contributed three reduction rules. For example, if two vertices $u$ and $v$ have more than $k$ common neighbors then the edge~$\{u,v\}$ has to be in the solution and is added if it is not present. Moreover, if $u$ and $v$ have more than $k$ non-common neighbors, \ie~vertices that are either neighbors of $u$ but not $v$ or vice versa,
then the edge $\{u,v\}$ does not belong to the solution. Lastly, if $u$ and $v$ have more than $k$ common and more than $k$ non-common neighbors, then the given instance has no solution. Overall, their method performs best using a refined branching method with re-kernelization. Interestingly, the experimental analysis of their algorithm shows that binary search may not be the best way to implement an fixed-parameter based approach for cluster editing.

Guo~\cite{guo2009more} later gave parameter-independent data reductions based on critical cliques, obtaining a linear kernel of $4k$ vertices, which was later improved by Chen and Meng~\cite{chen2012kernel} to $2k$. B\"ocker\etal~\cite{Boecker2011} introduced additional parameter-independent data reductions and find that preprocessing is possible if the number of edge modifications is significantly smaller than the number of vertices in the graph. In addition to the parameter-independent rules they combine their technique with the parameter-dependent reductions from above with lower and upper bounds. B\"ocker\etal~find that they can effectively reduce graphs that satisfy $k\leq 25|V|$, whereas the reductions due to Guo~\cite{guo2009more} are only effective for $k\leq |V|/2$. Their experiments show that computing exact solutions for cluster editing is no longer limited to small or almost transitive graphs. Afterwards, B\"ocker\etal~\cite{10.1007/978-3-540-85097-7_1,DBLP:conf/apbc/BockerBBT08} extended their results to the weighted version of the problem in which the weight of an edge yields the cost of deleting or inserting it, and the goal is to apply a set of edge modifications with minimum total weight. To this end, they include non-trivial extensions of the data reduction rules of the unweighted case. Additionally, they present a technique to merge vertices which drastically improves the running time of their algorithm. 
Recently, Bastos\etal~\cite{DBLP:journals/jco/BastosOPSMP16} combine exact methods with local search heuristics. 
More precisely, the authors propose a GRASP and an ILS metaheuristic with different neighborhoods as well as a new reduction rule for the problem. They show that the used data reduction rules can speed up linear programming for some instances up to 95\% decreased runtime after using reduction rules and 41\% decreased runtime on average on the instances that the solver could solve to optimality. 
\begin{openproblem}
Is it possible to compute small kernels in practice if the parameter $k$ is larger than $25|V|$? Are there any specific data reduction rules for that case? If an instance in practice does not reduce well, does that help to obtain bounds on the parameter~$k$?
\end{openproblem}

Since the parameter $k$ is often large compared to the number of vertices, fixed-parameter algorithms may not always be practical. There has been several attempts to use other parameters such as the number of missing edges per cluster as well as the number of edges between clusters~\cite{HeggernesLNPT10}, the total number of edge modifications per vertex \cite{abu17,KU12}. Abu-Khzam~\cite{abu17}, using local parameters that bound the amount of (either or both) edge addition and deletion per vertex resulted in a number of reduction rules, showed how to solve much larger problem instances and apply the problem effectively in data analysis \cite{Barr2020,Barr0AC19}.

\subsection{Multiterminal Cut} 
The \emph{multiterminal cut} problem with $k$ terminals is defined as follows:
Its input is an undirected edge-weighted graph $G=(V,E,w)$ with edge weights $w: E \mapsto \MdN_{>0}$ and its goal is to divide its set of vertices into~$b$ blocks such that each block contains exactly one terminal and the weight sum of the edges running between the~blocks~is~minimized. 
 It is a fundamental combinatorial optimization problem which was first formulated by Dahlhaus\etal~\cite{dahlhaus1994complexity} and Cunningham~\cite{cunningham1989optimal}. It is NP-hard for all $b \geq 3$~\cite{dahlhaus1994complexity}, even on planar graphs, and reduces to the minimum $s$-$t$-cut problem, which is in P, for $b=2$. The minimum $s$-$t$-cut problem aims to find the minimum cut in which the vertices $s$ and $t$ are in different blocks. Most algorithms for the minimum multiterminal cut problem use minimum s-t-cuts as a subroutine. Dahlhaus\etal~\cite{dahlhaus1994complexity} give a $2(1-1/b)$ approximation algorithm with polynomial running time. Their approximation algorithm uses the notion of \emph{isolating cuts}, \ie~the minimum cut separating a terminal from all other terminals. They prove that the union of the $b-1$ smallest isolating cuts yields a valid multiterminal cut with the desired approximation ratio. The currently best known approximation algorithm by Buchbinder\etal~\cite{buchbinder2013simplex} uses linear program relaxation to achieve an approximation~ratio~of~$1.323$.

Marx~\cite{marx2006parameterized} proves that the multiterminal cut problem is fixed-parameter tractable parameterized by multiterminal cut weight $\wgt(G)$.
Chen\etal~\cite{chen2009improved} give the first fixed-parameter algorithm with a running time of $4^{\wgt(G)}\cdot n^{\Oh{1}}$, later improved by Xiao~\cite{xiao2010simple} to $2^{\wgt(G)}\cdot n^{\Oh{1}}$ and by Cao\etal~\cite{cao20141} to $1.84^{\wgt(G)}\cdot n^{\Oh{1}}$. 

Recently, Henzinger\etal~\cite{DBLP:conf/alenex/HenzingerN020} engineer an algorithm that combines the branch-and-bound formulation of Xiao~\cite{xiao2010simple} with existing and new data reduction rules for the problem and present a shared-memory parallel branch-and-reduce algorithm for the multiterminal cut problem. Experiments indicate that this is orders of magnitude faster than previous ILP formulations for the problem that have been employed~by~practitioners. Later, reduction rules have been combined with local search algorithms for the problem~\cite{DBLP:journals/corr/abs-2004-11666}. The algorithm uses a wide variety of reduction rules with varying computational complexity; using vertex neighborhoods, edge connectivities, articulation points, maximum flows and more criteria to reduce the problem size; Henzinger\etal~\cite{DBLP:journals/corr/abs-2004-11666} report size reductions of up to multiple orders of magnitude in some instances, which make large instances solvable in practice. Additionally, they give an inexact algorithm that aggressively prunes subproblems which likely do not yield an improved solution.

\begin{openproblem}
Is there an efficient way to find semi-isolated small clusters that can be contracted (either exact or inexact contraction)? 
\end{openproblem}

\begin{openproblem}
The algorithm by Henzinger\etal~\cite{DBLP:conf/alenex/HenzingerN020} uses only reductions that guarantee that the optimal solution remains in the graph. Are there reductions that do not guarantee optimality but give good performance in practice?
\end{openproblem}

\section{Recent Advances for Problems in P}
\subsection{Minimum Cut}
Given an undirected graph with non-negative edge weights,
the \emph{minimum cut} problem is to partition the vertices into two sets so
that the sum of edge weights between the two sets is minimized. The size of a minimum cut is
often also referred to as the \textit{edge connectivity} of a
graph~\cite{henzinger2017flow,nagamochi1992computing}. 
Gomory and Hu~\cite{gomory1961multi}
observed that the (global) minimum cut can be computed with $n-1$ minimum
$s$-$t$-cut computations.
For the following decades, this result by Gomory and Hu was used to find better
algorithms for global minimum cut using improved maximum flow
algorithms~\cite{karger1996new}. 
Hao and Orlin~\cite{hao1992faster} adapt the push-relabel algorithm to pass
information to future flow computations. When a push-relabel iteration is finished,
they implicitly merge the source and sink to form a new sink and find a new
source. Vertex heights are maintained over multiple iterations of push-relabel. With
these techniques they achieve a total running time of $\Oh{mn\log{\frac{n^2}{m}}}$ for
a graph with $n$ vertices and $m$ edges, which is asymptotically equal to a
single run of the~push-relabel~algorithm.

However, for minimum cut algorithms to be viable for applications they must be fast on small data sets and scale to large data sets. Thus, an algorithm should have either linear or near-linear running time, or have an efficient parallelization. \emph{All} existing exact algorithms have non-linear running time~\cite{hao1992faster,henzinger2017flow,karger1996new}, where the fastest of these is the deterministic algorithm of Henzinger\etal~\cite{henzinger2017flow} with running time $\Oh{m \log^2{n} \log \log^2 n}$.
Although this is arguably near-linear theoretical running time, it is not known how the algorithm performs in practice. Even the randomized algorithm of Karger and Stein~\cite{karger1996new} which finds a minimum cut only with high probability, has $\Oh{n^2\log^3{n}}$ running time, although this was later improved by Karger~\cite{karger2000minimum} to $\Oh{m\log^3{n}}$ and recently improved further by Gawrychowski\etal~\cite{DBLP:conf/icalp/GawrychowskiMW20} to $\Oh{m\log^2{n}}$. 
The algorithm of Karger and Stein can be seen as probabilistic data reduction algorithms as they contract random edges to reduce the problem size, and give the correct answer with a certain probability.

Padberg and Rinaldi~\cite{padberg1990efficient} give a set of heuristics for
edge contraction. Chekuri\linebreak\etal~\cite{Chekuri:1997:ESM:314161.314315} give an
implementation of these heuristics that can be performed in time linear in the
graph size. Using these heuristics it is possible to sparsify a graph while
preserving at least one minimum cut in the graph. If their algorithm does not
find an edge to contract, it performs a maximum flow computation, giving the
algorithm worst case running time $\Oh{n^4}$. However, the heuristics can also be
used to improve the expected running time of other algorithms by applying them
on~interim~graphs~\cite{Chekuri:1997:ESM:314161.314315}.

\begin{openproblem}
Some reductions of Padberg and Rinaldi~\cite{padberg1990efficient} potentially check each triangle in a graph. Can pruning be used to check them efficiently?
\end{openproblem}

Nagamochi\etal~\cite{nagamochi1992computing,nagamochi1994implementing} give a minimum
cut algorithm which does not use any flow computations. Instead, their
algorithm uses maximum spanning forests to find a non-empty set of contractible
edges. 
The intuition behind the algorithm is as follows: imagine you have an unweighted graph with minimum cut  value exactly one. 
Then any spanning tree must contain at least one edge of each of the minimum cuts. 
Hence, after computing a spanning tree, every remaining edge can be contracted without losing the minimum cut.
Nagamochi, Ono and Ibaraki extend this idea to the case where the graph can have edges with positive weight as well as the case in which the minimum cut is bounded by $\hat \lambda$ and show how edges are identified using one modified breadth first search.
This contraction algorithm is run until the graph is contracted into a single
vertex. The algorithm has a running time of $\Oh{mn+n^2\log{n}}$. Stoer and Wagner~\cite{stoer1997simple} give a simpler variant of the algorithm of
Nagamochi, Ono and Ibaraki~\cite{nagamochi1994implementing}, which has a
the same asymptotic time complexity. The performance of this algorithm on
real-world instances, however, is significantly worse than the performance of the
algorithms of Nagamochi, Ono and Ibaraki or Hao and
Orlin, as shown in experiments conducted by J\"unger\etal~\cite{junger2000practical}.
Both the algorithms of Hao and Orlin, and Nagamochi, Ono and Ibaraki achieve close to linear running time on most benchmark instances~\cite{Chekuri:1997:ESM:314161.314315,junger2000practical}.

Based on the algorithm of
Nagamochi, Ono and Ibaraki, Matula~\cite{matula1993linear} gives a
$(2+\varepsilon)$-approximation algorithm for the minimum cut problem. The algorithm contracts more edges than the algorithm of
Nagamochi, Ono and Ibaraki to guarantee a linear time complexity while still
guaranteeing a $(2+\varepsilon)$-approximation factor. 
Inspired by random contractions, Henzinger\etal~\cite{DBLP:journals/jea/HenzingerNSS18} first gave an shared-memory parallel algorithm without guarantees on the cut size. The algorithm is randomized, and has running time $\Oh{n+m}$ when run sequentially. It repeatedly reduces of the input graph size with both heuristic and exact techniques, and then solve the smallest remaining problem with exact methods. The core idea of the inexact algorithm is that edges in densely connected regions (\ie~inside a cluster of a clustering) are unlikely to be in a minimum cut. The algorithm further uses exact reduction routines from Padberg and Rinaldi~\cite{padberg1990efficient}. 
For example, given a bound~$\hat\lambda$ on the minimum cut, one can obviously contract each edge having weight larger than~$\hat\lambda$, without loosing optimality.
Experimental results indicate that the algorithm finds optimal cuts
on almost all instances. At the same time, even when run sequentially, the algorithm is significantly faster (up to a factor of $4.85$) than other state-of-the-art algorithms.

Later, Henzinger\etal~\cite{DBLP:conf/ipps/HenzingerN019} engineered the fastest known \emph{exact} minimum cut algorithm for the problem. 
To do so, the authors incorporate the proposed inexact method, use better suited data structures and other optimizations as well as parallelization of exact methods.
More precisely, the exact algorithm uses the \emph{inexact} minimum cut algorithm from above~\cite{DBLP:journals/jea/HenzingerNSS18} to obtain a better approximate bound $\hat\lambda$ for the problem (recall that the algorithm almost always gave the correct result). As known reduction techniques depend on this bound, the better bound enables us to apply more reductions and to reduce the size of the graph much faster.
For example, edges whose incident vertices have a connectivity of at least $\hat\lambda$, can be contracted without the contraction affecting the minimum cut.
The new exact algorithm outperforms the state-of-the-art by a factor of up to $2.5$ already sequentially, and when run in parallel by a factor of up~to~$12.9$.
Similar reduction rules have later been used by Henzinger\etal~\cite{henzinger2020finding} to find all minimum cuts in graphs.

\subsection{Matching}
A matching $M$ of a graph $G=(V,E)$ is a subset of edges such that no two
elements of~$M$ have a common endpoint. Many applications require the
computation of matchings with certain properties, like being maximal (no edge can
be added to~$M$ without violating the matching property), having maximum
cardinality, or having maximum total weight $\sum_{e\in M}w(e)$, where $w$ is a positive weight function that assigns weights to edges. Although these
problems can be solved optimally in polynomial time, optimal algorithms are not
fast enough for many applications involving large graphs where we need near
linear time algorithms. For example, the most efficient algorithms for graph
partitioning rely on repeatedly contracting maximal matchings, often trying to
maximize some edge rating function $w$. We refer to Holtgrewe\etal~\cite{HSS10} for details
and examples. 
For the maximum cardinality matching problem, already in the 1980s data reduction rules have been proposed by  Karp and Sipser \cite{DBLP:conf/focs/KarpS81}. The rules are able to deal with vertices that have degree smaller than two. For example, it is quite easy to see that a vertex having degree zero can be removed from the graph, or if a vertex has degree one, then there is always a maximum matching that has this~edge~matched. 

M\"ohring and M\"uller-Hannemann~\cite{mohring1995cardinality} were among the first to use the rules to speed up heuristic algorithms for the general maximum cardinality problem. As exact algorithms for the matching problems typically search for augmenting paths, they can be speed up by using a good initial matching. Hence, later Langguth\etal~\cite{DBLP:journals/jea/LangguthMS10} analyzed the effects of various initializations on the total running time of several exact algorithms for the bipartite maximum cardinality problem and are able to achieve significant speed-ups. 

Korenwein\etal~\cite{DBLP:conf/esa/KorenweinNNZ18} implement (near-)linear time data reduction rules for the unweighted case as well as the positive-integer-weight case. Applied reductions include Karp-Sipser rules, as well as rules due to Mertizios\etal~\cite{DBLP:conf/mfcs/MertziosNN17} who have also shown that the maximum cardinality matching problem admits a kernel with at most $12k$ vertices and $13k$ edges where $k$ is the feedback edge number. Moreover, Koana\etal~transfer results from vertex cover to the matching problem, e.g. crown and LP-based data reductions. Experiments indicate that using data reduction rules can speed up state-of-the-art solvers by a factor of 4.7 for the unweighted case and 12.72 on average in the weighted case.
\begin{openproblem}
Can exhaustive application of reduction rules in \cite{DBLP:conf/esa/KorenweinNNZ18} be done in linear time?
Are there more rules that can be transferred from vertex cover to the matching problem that can be applied in near-linear time?
\end{openproblem}

Kaya\etal~\cite{DBLP:conf/alenex/KayaLPU20} also use Karp-Sipser based kernels for bipartite graph matching---in particular, the authors describe an efficient implementation as well as modifications to reduce time complexity on worst-case instances. Their implementation is about a factor 2 faster then the general purpose implementation of Koana\etal~\cite{DBLP:conf/esa/KorenweinNNZ18}.
Recently, Panagiotas and U{\c c}ar~\cite{panagiotas:hal-02463717} engineer fast almost optimal algorithms for bipartite graph matching. To this end, the authors investigate two randomized algorithms by Karp\etal~\cite{DBLP:journals/mor/KarpKV94} and Goel\etal~\cite{DBLP:conf/stoc/GoelKK10} and convert them to efficient heuristics for bipartite graphs. In particular, the algorithm by Karp~\cite{DBLP:journals/mor/KarpKV94} incorporates Karp-Sipser rules. Both of their heuristics run in near linear time and obtain matchings whose cardinality is more than 99\% of the maximum. 

\begin{openproblem}
Is it possible to implement the degree-2 vertex Karp-Sipser rule  in linear time?
\end{openproblem}

\section{Engineering Techniques}
Engineering techniques are necessary to make data reduction algorithms scale in practice. We give a short overview of techniques that are currently used in practice.  
The techniques we reference here include dependency checking, reduction tracking, plateau/increasing~data transformations, limiting to simple and fast reductions,  reduce and peeling, limited reductions, on the fly reductions and lastly parallelization.

\emph{Dependency checking} allows pruning of reductions when they will provably not succeed, therefore significantly reducing the number of failed reductions.
To compute a kernel, algorithms typically apply their
reductions~$r_1, \dots ,r_j$ by iterating over all reductions and trying to
apply the current reduction $r_i$ to all vertices. If $r_i$ reduces at
least one vertex, they restart with reduction~$r_1$. When reduction~$r_j$ 
is executed, but does not reduce any vertex, all reductions have been applied
exhaustively, and a kernel is found. Trying to apply every reduction to all
vertices can be expensive in later stages of the algorithm where 
few reductions succeed. The algorithm may repeatedly
attempt to apply the same reduction to a vertex even though
the graph has not changed sufficiently to allow the reduction to succeed. 
Checking dependencies between reductions~\cite{hespe2019scalable}, allows to
avoid applying certain local reductions when they
will provably not succeed, e.g. if their relevant neighborhood did not change since the reduction was last checked.  Therefore dependency checking keeps a
set $D$ of \emph{viable} candidate vertices: vertices whose relevant neighborhood has changed
and vertices that have never been considered for reductions. Then reductions are only applied at candidates that are in the set $D$. This avoids a lot of work and can speed up kernel computations significantly.

\emph{Reduction tracking.} The algorithm by Hespe\etal~\cite{hespe2019scalable}
stops local reductions when they are not effectively reducing the global graph sizes.
It is not \emph{always} ideal to apply reductions exhaustively---for example, if only few reductions will succeed and they are costly.
During later stages of a data reduction algorithm, local reductions may lead to very few graph changes.
Therefore, it may be better to stop local reductions early instead of performing them exhaustively and switch to global, more expensive reductions that may change the graph more significantly. 
Although the resulting graph is kernel-like, it may be possible to reduce it further. Such a graph is called a \emph{\quasikernel}.
Note, however, that this is a trade-off between size of the kernel and kernelization speed. 

\emph{Plateau/Increasing Transformations.} The general scheme in data reduction is to apply reductions exhaustively until non of the available reductions can be applied anymore. Gellner\etal~\cite{alex2020boosting} engineer new generalized data reduction and transformation rules for the weighted independent set problem. A key feature of this work are some transformation rules that can \emph{increase} the size of the input. Surprisingly, these so-called
\emph{increasing transformations} can simplify
the problem and also open up the reduction space to yield even smaller
irreducible graphs later throughout the algorithm. Overall, for the weighted independent set problem, this yields significant speed ups and enables the authors to solve more instances to optimality than previously possible. 

\emph{Simple Reductions.} 
Often the smallest kernels (or seemingly equivalently, the most varied reductions) give the best chance at finding solutions. For instance, the reductions used by Akiba and Iwata~\cite{akiba-tcs-2016} for the maximum independent set problem are the \emph{only} ones known to compute an exact solution on certain large-scale graphs, and these reductions are further successful in computing exact solutions in an evolutionary approach~\cite{redumis-2017}. However it is not always beneficial to compute the smallest kernel possible. Fast and simple reductions can compute kernels that are ``small enough'' for local search to quickly find high-quality, and even exact, solutions much faster than the reductions used to find the smallest kernels~\cite{chang2017,dahlum2016accelerating}. Fast and simple reductions can even be used to solve many large-scale instances exactly~\cite{strash2016power} just as quickly as the algorithm by Akiba and Iwata~\cite{akiba-tcs-2016}.

\emph{Reduce and Peeling.} 
Lamm\etal~\cite{redumis-2017} showed that including reductions in a branch-and-reduce inspired evolutionary algorithm for the independent set problem enables finding exact solutions much faster than provably exact algorithms. To this end, reductions are applied exhaustively. Once a kernel is computed, vertices that are unlikely to be in the solution, e.g. vertices having a very large degree, are removed from the graph and hence excluded from the solution. The algorithm then proceeds recursively.
Chang\etal~\cite{chang2017} improved on this result by implementing reduction rules to reduce the lead time for kernelization for local search. 
They introduce ``reducing--peeling'' to find a large initial solution for local search. This technique can be viewed as computing one path through the search space of a branch-and-reduce algorithm: they repeatedly exclude high-degree vertices and kernelize the graph until it is empty, then they take the solution found as an initial solution for~local~search.

\emph{Limited Reductions.} Sometimes reductions can be very expensive, for example if their running time depends on the number of edges in 
the neighborhood of a certain vertex.
However, as mentioned above it is often not necessary to compute the smallest possible kernel in practice. Hence, a common technique in practice is to exclude such reductions, for example, if the degree of a vertex is too large. An application of this technique is due to Ost\etal~\cite{ost2020engineering} for the vertex ordering problem, where the simplicial vertex reduction rule is limited to vertices of degree at most 18.

\emph{On the Fly Reductions.}
Kernelization can be used as a preprocessing step to other
algorithms. However, reductions are also used to reduce the size of the search space of local search algorithms without losing solution quality.
Dahlum\etal~\cite{DBLP:conf/wea/DahlumLS0SW16} apply a set of simple reductions on
the fly for the independent set problem. For this algorithm, they only use reductions that do not require the graph to be modified.
For example, some vertices, e.g. one can just mark isolated vertices. This speeds up local search significantly.

\emph{Parallelization.}
A general technique to speed up algorithms is parallelization. Also in data reduction parallelization is used to speed up preprocessing times. 
For example, ``local'' reduction rules have been parallelized by using for graph partitioning techniques, \ie~each process works on a subgraph and applied reductions only in his subgraph~\cite{hespe2019scalable}.
At the same time, there are also attempts~\cite{hespe2019scalable} to parallelize more expensive ``global'' reductions, \eg~reductions that need to access the whole input instance. 

\emph{Targeted Branching.} Branch-and-reduce algorithms often use of vertex selection strategies that are carried over from existing branch-and-bound approaches. However, these selection strategies often do not take into account that removing certain vertices from the graph might result in an increase of the reduction space, which in turn might lead to smaller search trees. Gao\etal~\cite{gao2018exact} thus present a dynamic vertex selection strategy that also takes into account one of their reduction rules and uses a degree-based selection as a fallback. Their experiments indicate that this strategy is able to provide better results when compared to a purely degree-based~selection~rule.

\emph{Data-Driven Reductions.} Eblen\etal~\cite{eblen2012maximum} show the benefits of using application-specific reduction rules that exploit prior knowledge of the input space. In particular, they use a reduction rule that is based on the empirical evaluation of large transcriptomic graphs and is able to drastically reduce the running time of their algorithm on similar instances. However, this comes at the drawback of a decrease in performance for random graphs.

\section{Open Problems and Future Work}
We already discussed problem specific open problems throughout this article. Here, we list some general open questions that apply to a range of problems touched in this survey.
For example, in a branch-reduce algorithm can we branch to specifically get graphs that reduce better using the available portfolio of reductions? As a concrete example, as stated above, it may be helpful to end up with a lot of independent connected components and to achieve this one may be able to branch on a small vertex separator first. 
For most of the problems it is currently unknown, what makes problems hard to kernelize, \eg~when does which reduction work well in practice and why? From the theory perspective of a practitioner, it would be better to have an analysis of the expected kernel size, rather than the worst case so as to get more realistic results in practice. One does not always need a single optimal solution, but a diverse set of high-quality solutions. Theoretical approach for this have been proposed~\cite{DBLP:conf/ijcai/BasteFJMOPR20}, however, they still remain untested in practice. Probabilistic reductions have not yet been tried in practice. On the other hand, most of the dynamic techniques that maintain a problem kernel have also not yet been implemented. A problem that needs careful investigation is the order in which reduction rules are applied, \eg~when is it good to apply which reduction rule first?  Lastly, consider an instance for a problem on which you already applied all data reduction rules at hand exhaustively. Moreover, assume that you already have an optimal solution on the reduced instance. Is it possible to discover new rules by applying machine learning learning techniques on such instances?

\begin{acknowledgement}
Partially supported by DFG grants MN 59/1-1 and SCHU 2567/1-2.
\end{acknowledgement}